\documentclass[9pt,twocolumn,twoside]{opticajnl}
\usepackage{graphicx,amsmath,relsize,color,mathtools,bm,float}
\journal{josaa} % Choose journal (ao,jocn,josaa,josab,ol,optica,pr)

\setboolean{shortarticle}{false}
\title{Anticaustics in a Fabry-Perot interferometer}

\author[1,2,*]{Luis L. S\'anchez-Soto}
\author[1]{Juan J. Monz\'on}
\author[2,3,4]{Gerd Leuchs}

\affil[1]{Departamento de \'Optica, Facultad de F\'{\i}sica, Universidad Complutense, 28040~Madrid, Spain}

\affil[2]{Max-Planck-Institut f\"ur die Physik des Lichts,  91058~Erlangen, Germany} 

\affil[3]{Institute of Optics, Information and Photonics, University Erlangen-Nuremberg, 91058 Erlangen, Germany}

\affil[4]{Institute of Applied Physics, Russian Academy of Sciences, 603950 Nizhny Novgorod, Russia}

\affil[*]{Corresponding author: lsanchez@fis.ucm.es}

\begin{abstract}
We address the response of a Fabry-Perot interferometer to a monochromatic point source. We calculate the anticaustics (that is, the virtual wavefronts of null path difference) resulting from the successive internal {reflections} occurring in the system. They turn to be a family of ellipsoids {(or hyperboloids)} of revolution, which allows us to reinterpret the operation of the Fabry-Perot from a geometrical point of view that facilitates comparison with other apparently disparate arrangements, such as Young's double slit.
\end{abstract}

\setboolean{displaycopyright}{true}

\begin{document}

\maketitle

\section{Introduction}

Young’s double slit experiment~\cite{Young:1804wh} is the epitome of the phenomenon of interference and it is still found in any modern optics textbook~\cite{Born:1999yq}. Initially, the experiment was a crucial step in establishing the wave theory of light. Ironically, it played a quintessential role in ellaborating the concept of {wave-particle} duality, the ``central mystery" of quantum theory~\cite{Feynman:2011aa}. 

The standard interpretation of Young's experiment  is based on the superposition of the wavefronts emanating from two coherent point sources. Similar arrangements have also been developed during the XIXth century, including Fresnel's biprism~\cite{Fresnel:1866ux}, Lloyd's mirrors~\cite{Lloyd:1831ti} and Billet's split lens~\cite{Billet:1858uh}, among others. In all of  these cases, two  point sources (real or virtual) are required to explain the observed intensity pattern.   

One might rightly wonder how this picture is modified when many mutually coherent waves are superposed. The Fabry-Perot {(FP)} interferometer is the paradigm of this situation. Apart from its luminosity, the distinctive feature of this setup is its narrow resonances, which is the basis for its extensive use in high-resolution spectroscopy, interferometry, and laser resonators~\cite{Hernandez:1986zr,Vaughan:1989qn}. However, the ordinary approach assumes plane waves; that is, the source is at infinity.  Therefore, it seems clear that to appreciate the analogies and differences between two- and {multiple}-source interference, one should look at the response of the FP to a point source.   

In this paper, we take this unorthodox viewpoint. To this end, we analyze the wavefronts generated at each of the multiple reflections occurring in the FP. Although methods for obtaining the local properties of the refracted wavefronts are well established, the global properties are more elusive. The reason is that, in general, wavefronts may be very intricate surfaces, particularly in the neighborhood of a focus, and their analytic expressions  might become quite involved~\cite{Stavroudis:1972wv}. 

However, it is often possible to extract from a family of wavefronts one that is simpler in form, whose global properties are more comprehensible, but which, nonetheless, is representative of the entire family.  It corresponds to the {privileged} wavefront of zero algebraic optical path difference. This concept has been reinvented under a variety of names; {an indicative list of dates where it appears with different designations has been compiled by Ouellette~\cite{Ouellette:2018ww}: anticaustic (Bernouilli, 1692~\cite{Bernouilli:1999ul}), secondary caustic (Quetelet, 1829~\cite{Quetelet:1829ts}), orthogonal trajectory (Cayley, 1857~\cite{Cayley:1857tb}, simplest orthotomic curve (Herman, 1900~\cite{Herman:1900vb}), zero phase-front (Eaton, 1952~\cite{Eaton:1952ux}), emerging wavefront of null optical path (Damien, 1952~\cite{Damien:1955te}), archetypical wavefront (Stavroudis, 1969~\cite{Stavroudis:1969ux}), zero-distance phase front (Cornbleet, 1984~\cite{Cornbleet:1984wh,Gitin:2015um}), and phase front (Avenda\~{n}o-Alejo \emph{et al}, 2015~\cite{Avendano-Alejo:2015vz}).}

{In a comprehensive historical review, Chastang and Farouki~\cite{Chastang:1992ur} endorse the use of anticaustic, returning to the  original Bernouilli's suggestion.  This  was strongly supported by the late Emil Wolf~\cite{Wolf:te}.} We believe that reintroducing the term is appropriate, and constitutes a tribute to a great geometer and to a great physicist.  

The anticaustic has been explicitly calculated in a few simple cases, including refraction by a plane (ellipse/hyperbola) and refraction (Cartesian ovals) and reflection (lima\c{c}on of Pascal) by a sphere~\cite{Ouellette:2018ww,Farouki:2022uq}. Albeit the concept has been extensively used in microwaves~\cite{Cornbleet:1994vg}, it has not received the attention it deserves in optics. We hope that this neglect will be repaired with our discussion of the FP and the notion of anticaustic will experience a merited revival, given its {beauty and} potential usefulness.

\section{Anticaustics in a Fabry-Perot}

\begin{figure}[t]
\centering 
\includegraphics[width=.80 \columnwidth]{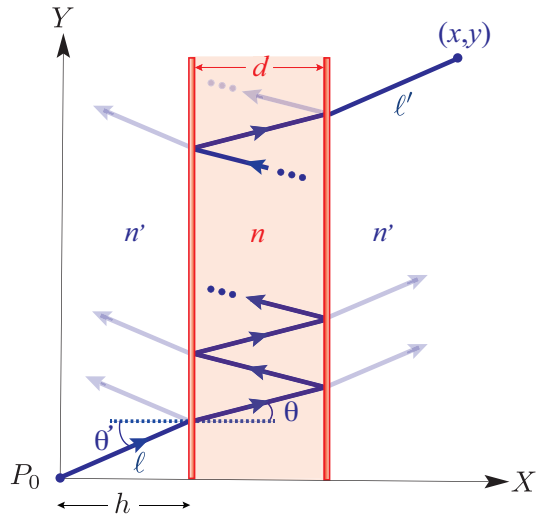}
\caption{Scheme of the FP. The system is illuminated by a monochromatic point source $P_0$ located at the origin {of the coordinate system}. A typical ray defining the anticaustic is marked in bold blue, while the faint lines indicate other rays generated in the multiple reflections.}
\label{fig:setup}
\end{figure}

We will examine a simplified  model of a FP interferometer, which consists of a plane parallel transparent plate of refractive index $n$ and thickness $d$ surrounded by a medium of refractive index $n^{\prime}$. As heralded in the Introduction, we consider that the system is illuminated by the spherical wavefronts arising from an ideal point source $P_{0}$ at a distance $h$ of the first interface. {We take a coordinate system with origin at the point $P_{0}$} and with the $X$ axis normal to the interfaces, as sketched in Fig.~\ref{fig:setup}. Without loss of generality, the problem will be treated as two dimensional. We will only look at the transmitted pattern, although an analogous treatment can be done for the reflected one.  

%%%%%%%%%%%%%%%%%%%%%%%%%%%%%%%%%%%%%%%%%%%% 
%%%%%%%%%%%%%%%%%%%%%%%%%%%%%%%%%%%%%%%%%%%%%%

The rays proceeding from the point source $P_0$ form a homocentric pencil, so they constitute a normal congruence since every ray of the pencil is cut orthogonally by spheres centered on the mutual point of intersection of the rays. After being refracted, the resulting pencil (in general, no longer homocentric) will again form a normal congruence. This is the basis of the Malus-Dupin theorem~\cite{Born:1999yq}, which guarantees the existence of wavefronts at the exit of the interferometer.   

Consider a typical ray from $P_0$ striking the first interface at an angle of incidence $\theta^{\prime}$ after traveling a distance $\ell$. Inside the plate, it experiences multiple reflections: after $2k+1$ {(with $k=0, 1, \ldots$)} of these reflections, the ray is transmitted, as indicated in the figure. Finally, let the transmitted ray travel a distance $\ell^{\prime}$ in the second medium (of index $n^{\prime}$). 

By direct inspection, one can check that the coordinates of the endpoint $(x, y)$ of this ray can be expressed as
\begin{equation}
\begin{split}
\label{eq:coord1}
x & = h + d + \ell^{\prime} \cos \theta^{\prime} \, ,  \\
y & = h \tan \theta^{\prime} + (2k+1) d \tan \theta + \ell^{\prime} \sin \theta^{\prime}  \, , 
\end{split}
\end{equation}
where $\theta$ is the angle of refraction in the medium $n$, which is related to the angle of incidence according to Snell’s law $n^{\prime} \sin \theta^{\prime} = n \sin \theta$. 

The optical path length of the considered ray, denoted by $\Delta$ and computed from the source $P_{0}$, is
\begin{equation}
\label{eq:OPL1}
\Delta (x,y) = n^{\prime} \ell + (2k+1) \frac{n d}{\cos \theta} + n^{\prime} \ell^{\prime} .
\end{equation}
To ensure that $(x, y)$ define a wavefront, we have to impose that $\Delta (x,y)$ takes a constant value, much in the spirit of Malus-Dupin. This gives a one-parameter family of wavefronts which constitutes the image of the  family of object wavefronts centered at $P_{0}$. 

%%%%%%%%%%%%%%%%%%%%%%%%%%%%%%%%%%%%%%%%%%%% 
\begin{figure}[t]
\centering 
\includegraphics[width= 0.95\columnwidth]{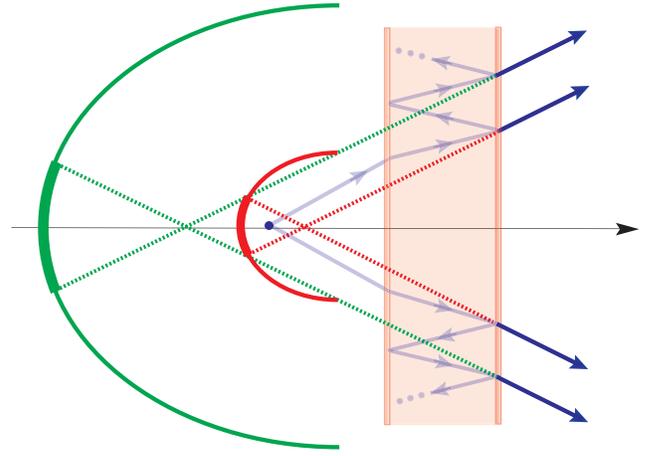}
\caption{First (red) and second (green) anticaustic in a FP interferometer. The broken lines represent virtual rays emanating from the anticaustics.}
\label{fig:waves}
\end{figure} 
%%%%%%%%%%%%%%%%%%%%%%%%%%%%%%%%%%%%%%%%%%%

Out of this family, we extract a single wavefront, the anticaustic, for which the optical path length $\Delta$ in \eqref{eq:OPL1} is zero: this fixes at once
\begin{equation}
\ell^{\prime} = - \frac{h}{\cos \theta^{\prime}} - (2k+1)  \frac{\mathfrak{n} d}{\cos \theta}  \, ,
\end{equation} 
where, for simplicity, we have introduced the relative index $\mathfrak{n} = n/n^{\prime}$ and we have taken into account that $\ell = h/\cos \theta^{\prime}$. {The negative sign in the optical} path length indicates that we are dealing with a virtual ray, marked with broken lines in Fig.~\ref{fig:waves}. 

Replacing this value {of $\ell^{\prime}$} in \eqref{eq:coord1} we get
\begin{equation}
\label{eq:wf1}
\begin{split}
{X} & = d - (2k+1) \mathfrak{n} d  \frac{\cos \theta^{\prime}}{\cos \theta} \, , 
\\
{Y}  & = (2k+1) d \tan \theta  - (2k+1) \mathfrak{n} d \frac{\sin \theta^{\prime}}{\cos \theta}    \, ,
\end{split}
\end{equation}
{where we denote by $(X, Y)$ the coordinates of the anticaustic point to distinguish them from the coordinates of the endpoint of the ray.}

%%%%%%%%%%%%%%%%%%%%%%%%%%%%%%%%%%%%%%%%%%%% 
\begin{figure*}[t]
\centering 
\includegraphics[width=1.60 \columnwidth]{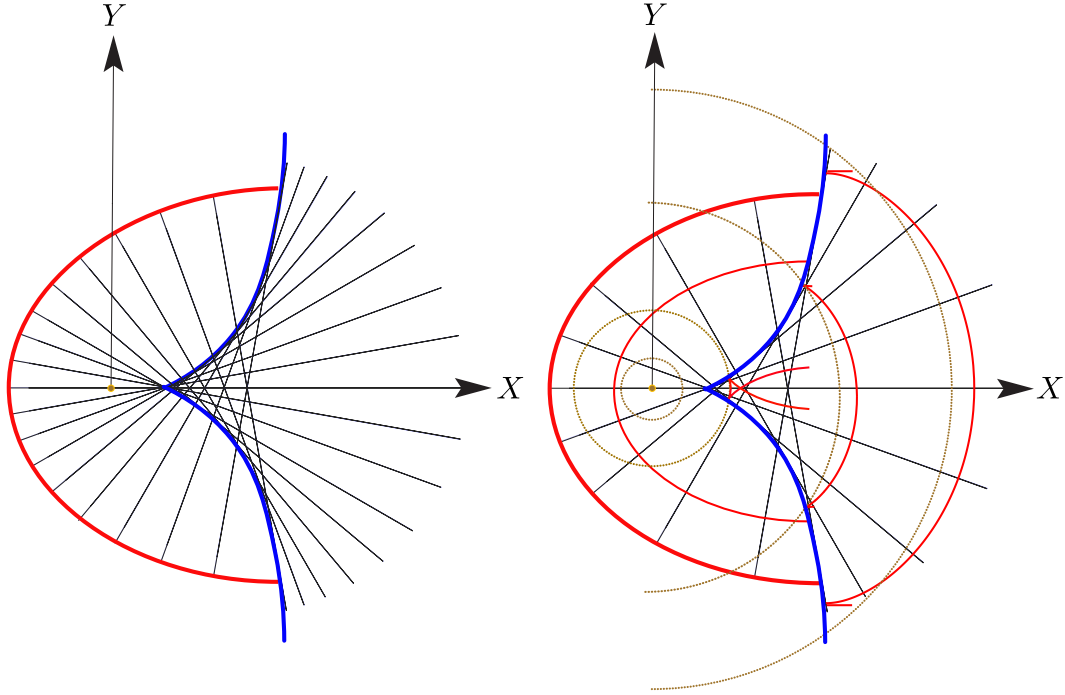}
\caption{{(Left) First anticaustic of the FP as in Fig.~\ref{fig:waves}, showing the corresponding caustic, which is  the envelope of the refracted rays. We have used $\mathfrak{n} = 1.3$ and $d=2$ (in arbitrary units). (Right) Propagation of the previous anticaustic at four different times. We have  included spherical wavefronts (dotted circles), corresponding to a single point source located at the position $P_0$.}}
\label{fig:caust}
\end{figure*} 
%%%%%%%%%%%%%%%%%%%%%%%%%%%%%%%%%%%%%%%%%%%%%% 

Finally, we use Snell's law  to eliminate $\theta^{\prime}$; the result reads
\begin{equation}
\label{eq:wf1}
\begin{split}
{X} & = d - (2k+1) \mathfrak{n} d \sqrt{1 + (1 - \mathfrak{n}^{2}) \tan^{2} \theta} \, , 
\\
{Y}  & = (2k+1) ( 1 - \mathfrak{n}^{2}) d \tan \theta     \, . 
\end{split}
\end{equation}
This is the parametric form of the anticaustic we were looking for~\cite{Martinez-Risco:1976tj}. To make things crystal clear, the explicit equation can be easily obtained by eliminating $\theta$ between these two equations:  
\begin{equation}
\label{eq:wf2}
\frac{( {X} - d)^{2}}{(2k+1)^{2} \mathfrak{n}^{2}d^{2}} +   
\frac{{Y}^{2}}{(2k+1)^{2}  (\mathfrak{n}^{2} -1) d^{2}} = 1 \, . 
\end{equation}
This represents a series of ellipses  for the usual case $\mathfrak{n} > 1$ ($n^{\prime}  < n$), whereas  for  $\mathfrak{n} < 1$ ($n^{\prime}  > n$) they are hyperbolas. These wavefronts are unique and provide exact information about every ray emanating from $P_{0}$ and refracted by the FP.  

\section{Discussion}

The action of the FP can be seen as producing the superposition of this set of ellipses (or hyperbolas) labeled by the integer $k$. {In what follows, we concentrate on the case of elliptical wavefronts, although the results can be directly translated to the hyperbolic ones.} All these wavefronts are concentric and have common axes of directions, the major ones being coincident with the $X$ axis. The center of all of them is {at the point $(d,0)$ and thus, at a distance $d$ of the source $P_0$}.  The respective semiaxes are given by 
\begin{equation}
\begin{split}
{a_{k}} & = (2k+1) \mathfrak{n} d \, , \\
{b_{k}} & = (2k+1) \sqrt{\mathfrak{n}^{2} - 1} \;  d\, .
\end{split}
\end{equation}
{The distance from each focus to the center is $f_{k}= \sqrt{a_{k}^{2} - b_{k}^{2}} = (2k+1) d$. The eccentricity of all these ellipses is  the same; viz, $\varepsilon = f_{k}/a_{k} = 1/\mathfrak{n} < 1$. Note that these elliptical wavefronts are independent of the distance $h$, so they do not change in shape nor position when the plate undergoes a translation; showing a remarkable invariance.}

In Fig.~\ref{fig:waves} we have plotted the first two anticaustics obtained by a simple transmission ($k=0$) and a transmission and two internal reflections ($k=1$). The physical meaning of these surfaces is apparent from the figure. {In practice, only a certain cap is effective in  each ellipse. In Fig.~\ref{fig:waves} we have marked these caps for the plotted rays. The larger the plate, the greater the extension of the cap. The  limit of the effective region can be found  by looking at the ray that emerges forming a greater angle with the normal to the interfaces; when virtually prolonged, it gives, on the corresponding ellipse, the point that serves as the limit.}

Consider, for example, the case $k=0$. Suppose a spherical wavefront originates from $P_{0}$ at $t=0$ and, after refractions, takes the form $W$ at time $t$. By propagating $W$ backward in time in the medium $n^{\prime}$ we get an initial wavefront $W_{0}$ at $t=0$. \emph{Mutatis mutandis}, the propagation of $W_{0}$  (without any refraction) yields the true wavefront $W$ at time $t$.

As stressed in the Introduction, the wavefront $W_{0}$ was investigated by Bernouilli. The term anticaustic, he concocted, has a mathematical origin: {actually the evolute of $W_{0}$ (i.e., the locus of all its centers of curvature) is precisely the caustic. Conversely, the wavefront $W_{0}$ is an involute of the caustic~\cite{Farouki:2001vh}.}

%%%%%%%%%%%%%%%%%%%%%%%%%%%%%%%%%%%%%%%%%%%% 
\begin{figure}[t]
\centering 
\includegraphics[width=.85 \columnwidth]{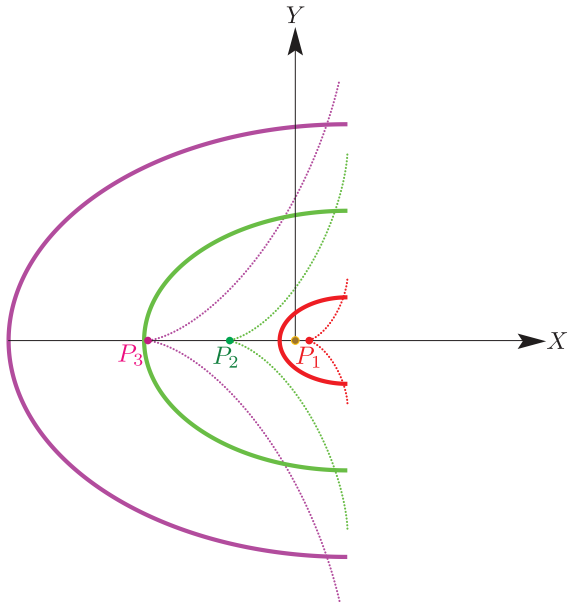}
\caption{{First three anticaustics of the same FP as in Fig.~\ref{fig:caust}, with their corresponding caustics. The source $P_0$ (not labeled) is located at the origin of the coordinate system. In the paraxial regime, the  equivalent virtual sources $P_{1}, P_{2}, P_{3}$  appear located at the cusps of the caustics and they are separated by $2d/\mathfrak{n}$.}}
\label{fig:points}
\end{figure} 
%%%%%%%%%%%%%%%%%%%%%%%%%%%%%%%%%%%%%%%%%%%%%%

{In consequence, the evolutes of \eqref{eq:wf2} are the caustics generated in the successive internal reflections. The calculation is straightforward, with the result~\cite{Gray:1993aa}}
\begin{equation}
{\mathfrak{n}^{2/3} (X-d)^{2/3} + (\mathfrak{n}^{2}-1)^{1/3} Y^{2/3} = (2k+1)^{2/3} \, d^{2/3} \,. }
\label{eq:astr}
\end{equation}
{This constitutes a series of astroids, all centered at $(d,0)$, in turn the common center of the ellipses. The amazing properties of these curves have been discussed in detail in the literature~\cite{Lockwood:1963aa,Yates:2012aa}. In Fig.~\ref{fig:caust} we plot the first anticaustic ($k=0$) and its corresponding caustic. Although the astroid is a 4-cusped curve, we concentrate on the horizontal left cusp; they  appear at positions}
\begin{equation}
{P_{k} = d - (2k+1) \frac{d}{\mathfrak{n}} \, ,}
\end{equation}
{as one can check directly from \eqref{eq:astr}. They are  thus separated by a distance $2d/\mathfrak{n}$, as it shown in Fig.~\ref{fig:points}.} 

{Once the anticaustic $W_{0}$ is known, the propagated wavefront $W$ is equidistant from (or offset from) $W_{0}$. To obtain the equidistant curve, we consider the (oriented) normals of $W_{0}$:  the equidistant curve is formed by the points at a distant $vt$ (with $v$ the velocity of light in the medium) in the forward direction from the points of the initial curve along each normal. This is represented in Fig.~\ref{fig:caust}.  For small $t$ the equidistant curve is smooth. But from some value of $t$ on (namely, this critical value being the minimal curvature radius of the curve), the equidistant curve acquire{}s singularities. They were already studied by Cayley~\cite{Arnold:1991vh}, and they are semiclassical cusped edges and swallowtails. They can be appreciated in the wavefronts appearing inside the caustic. In physical terms, this means that the successive images by the interfaces of the point $P_{0}$ are no single points, but the system presents aberrations. This is also apparent in Fig.~\ref{fig:points}.}

{In Fig.~\ref{fig:caust} we have also included the spherical wavefronts associated to the source $P_{0}$. We can see significant differences with the propagated anticaustics. For small aperture angles $\theta^{\prime}$, we are in the paraxial regime and the effective caps in the ellipses can be considered locally as spheres. In this case, the FP appears as the interference of the virtual point sources $P_{1}$, $P_{2}, \ldots$, separated by a distance  $2d/\mathfrak{n}$~\cite{Sanchez-Soto:2016vy}. In Fig.~\ref{fig:points} we can appreciate that these virtual sources are precisely at the cusps of the caustics previously discussed. This makes clear contact with Young's experiment and bears a close resemblance with a diffraction grating. There is, however, one important difference: in the FP the amplitudes of each equivalent virtual source decrease due to the reflection and transmission coefficients in each interface.}  

Alternatively, these secondary sources can be seen as a far-field uniform linear array antenna, with excitation amplitudes arranged according to a geometric sequence.  This configuration can be addressed with the standard methods of antenna theory~\cite{Balanis:2016aa}.   

Since these secondary sources get fainter farther away from the real source, only an effective number of them contribute. This number will depend on the reflectivity of the interfaces.  

As a final curiosity, we quote that this picture in terms of anticaustics gives an intuition about what happens when the interfaces are not strictly parallel, a conundrum in this field~\cite{Born:1999yq}. Now, the points $P_{0}$ and the successive images, $P_{1}, P_{2}, \ldots$ are spaced around a circle and the resulting diffraction grating  gives a high-order Bessel function~\cite{Mcgloin:2005aa}, which has associated oscillations on one side of the peak.

\section{Concluding remarks}

In summary, we have made extensive use of the notion of anticaustic to provide an alternative view of the response of an FP interferometer. The intriguing mathematical properties of these surfaces~\cite{Farouki:2022uq} confirm their relevance in optical problems. 

We stress that the benefit of this approach lies not in any inherent advantage in terms of efficiency in solving practical matters.  Rather, we expect that the beautiful geometrical picture presented here may be helpful in updating the modern views on the operation of such a relevant setup.

\begin{backmatter}
\bmsection{Funding} 
Financial support from the Spanish Ministerio de Ciencia e Innovaci\'on (Grant PGC2018-099183-B-I00) is gratefully acknowledged.

\bmsection{Acknowledgments} We would like to {acknowledge} Emil Wolf for bringing our attention to the concept of anticaustic and the historical details about the names under which this {notion} has been badged in the literature. 
{We thank two anonymous reviewers for their detailed and thoughtful comments that have helped us to improve this manuscript.}

\bmsection{Disclosures} The authors declare no conflicts of interest.

\bmsection{Data Availability Statement}  No data were generated or analyzed in the presented research.

\end{backmatter}

% Bibliography
%\bibliography{Fabry}

\end{document}